\documentclass [12pt,preprint]{aastex}
\input epsf
\journalid{}{}
\articleid{}{}
\begin{document}
\def\lax    {\ifmmode{_<\atop^{\sim}}\else{${_<\atop^{\sim}}$}\fi}
\def\gax    {\ifmmode{_>\atop^{\sim}}\else{${_>\atop^{\sim}}$}\fi}
\def\gtorder{\mathrel{\raise.3ex\hbox{$>$}\mkern-14mu
             \lower0.6ex\hbox{$\sim$}}}
\def\ltorder{\mathrel{\raise.3ex\hbox{$<$}\mkern-14mu
             \lower0.6ex\hbox{$\sim$}}}

\title{Observational Signatures of Black Holes: Spectral and Temporal
Features of XTE J1550-564}
\author{Lev Titarchuk\altaffilmark{3,1}, \&
C.R. Shrader\altaffilmark{1,2}}

\altaffiltext{1}{Laboratory for High--Energy Astrophysics,
NASA Goddard Space Flight Center, Greenbelt, MD 20771, USA;
shrader@grossc.gsfc.nasa.gov }
\altaffiltext{2}{Universities Space Research Association, Lanham MD}
\altaffiltext{3}{George Mason University/CEOSR; lev@lheapop.gsfc.nasa.gov;
lev@xip.nrl.navy.mil}

\vspace{0.1in}

\begin{abstract}
The theoretical predictions of the converging inflow, or Bulk-Motion
Comptonization model are discussed and some predictions are
compared to X- and gamma-ray observations of the high$-$soft state of
Galactic black hole candidate XTE J1550+564.
 The $\sim 10^{2}$-Hz QPO phenomenon tends to be detected in the
high-state at times when the bolometric luminosity surges and the
hard-power-law spectral component is dominant. Furthermore, the power in these features
increases with energy. We offer interpretation of  this phenomenon, as
oscillations of the innermost part of the accretion disk, which in turn
supplies the seed photons for the converging inflow where the hard power-law is formed
through Bulk Motion Comptonization (BMC).
We further argue that the noted lack of coherence between intensity variations of the
high-soft-state low and high energy bands is a natural consequence of our
model, and that a natural explanation for the observed hard and soft lag
phenomenon is offered.
In addition, we address some criticisms of the BMC model supporting our
claims with observational results.
\end{abstract}

\section{Introduction}

It has long been known that accreting stellar-mass black holes (BH) in
Galactic binaries exhibit a ``bi-modal" spectral behavior - namely the so
called high-soft and low-hard spectral states; see Liang (1998) for a recent
review. High and low in this context refer to the relative 2-10 keV luminosities
in a given system, which is related to the rate at which mass accretion is
taking place. An increase in the soft blackbody luminosity component leads to the
appearance of the extended power law. An important observational fact is
that this effect is seen as a persistent phenomenon only in BHCs,
and thus it is apparently a {\it unique} black hole signature.

Although in Neutron star (NS) systems  similar power law components are
detected in the intermediate stages (Strickman, Barret 1999; Iaria et al. 2000;
Di Salvo et al. 2000), they are of a  transient nature, disappearing with
the luminosity increase (Di Salvo et al.).
The low-hard state spectral form on the other hand, has been seen in
BH and NS binaries as well under certain condition. It was previously
believed that the hard component in NS systems was associated with the  low
luminosity states of bursters (Barret \& Grindlay 1995; Heindl \& Smith 1998) when
the soft component is suppressed.

It thus seems a reasonable assumption that the unique spectral signature
of the soft state of BH binaries is directly tied to the black hole event
horizon. This is the primary motivation for the Bulk Motion Comptonization Model
(BMC) introduced in several previous papers, and recently applied with striking
success to a substantial body of observational data. A complete theory of
black hole accretion must, however, be also able to accommodate in a
natural manner a growing number of empirical traits exhibited in the temporal
domain. For example, it is now well established that black hole X-ray binaries
exhibit quasi-periodic oscillation (QPO) phenomena in essentially three distinct
frequency domains; low, $\sim 0.1$ Hz, intermediate ($\sim 1-10$ Hz) and
high ($\sim 10^2$ Hz). It is likely that these are interrelated by some as
yet unspecified underlying physics. The high-frequency QPOs seem to occur
during periods of flaring, and when the spectra (although in the high-soft state)
tend to be relatively hard, i.e. the proportional hard-power- law
to thermal excess flux ratio is larger than usual. Furthermore, the QPO
amplitudes increase with energy, that is there is a higher degree of
modulation of the signal in the hard-power law than in the thermal excess component.
In addition to the QPO phenomena, it has been noted by Nowak et al
(2000) (also see Cui, Zhang, \& Chen 2000) that measures of the coherence
between the intensity variations in the hard power law and thermal
components is negligible.

In this paper we present empirical evidence in both the temporal and
spectral domains in support of our ideas on the converging inflow
phenomenon and the BMC model. In section 2, we describe the main observational
consequences of the Comptonization model for high-soft and low-hard states  based on the
presence of thermal and advection dominated (converging  inflow) motions
around the compact objects. We consider some empirical properties of accreting black
holes in Section 3, highlighted by recent observations of XTE~J1550-564. Interpretation
within the context of our BMC model is presented.
In section 4 discussion of issues such as the reliability of obtaining
the correct spectra from the deconvolution of time-varying, background dominated signals
is discussed. We also investigate possible high-soft-state features such as Compton
reflection and relativistically broadened line reported in the recent literature, and
present physical parameter estimation (distance-to-mass ratio) for
XTE~J1550-56. In addition we consider the interrelation between spectral and temporal black hole
properties, identifying further support of the converging-inflow hypothesis.
Summary and conclusions are offered in section 5.

\section{The Bulk-Motion Comptonization Model}

\subsection{An Overview and Synopsis of the Recent Literature}

As noted, the specific spectral and timing features of the X-ray
radiation characterizing the high-soft and low-hard states presumably
provide insights into the underlying physical processes. The
thermal Comptonization model (Sunyaev \& Titarchuk 1980; Titarchuk
1994; Hua \& Titarchuk 1995; Poutanen \& Svensson 1996) quite well fits the low-hard
state spectra with a typical electron temperature of $T_e\sim$60~keV and optical depth
$\tau\sim$1(Zdziarski 2000). These parameters seem to describe both galactic and
extragalactic  BH sources over the broad range of luminosities, characteristics
of those objects; the energy spectral indices $\alpha$ and electron
temperatures $kT_e$ are consistently about 0.7 and 60 keV respectively.
When the luminosity of the thermally radiating disk reaches
a certain threshold, the  power law steepens with the canonical value of
$\alpha\approx 0.7$ changing to $\alpha\approx 1.5$. Laurent \&
Titarchuk (1999, see also Titarchuk, Lapidus \& Muslimov 1998, hereafter
TLM98) argued that this type of the spectral phase transition is a consequence of
the bulk and thermal motions in a Compton cloud surrounding the black
hole.

Sunyaev \& Titarchuk (1978; 1980 hereafter ST80) first demonstrated, using
an analytical Comptonization theory, that the observed spectrum of Cyg X-1
(Sunyaev \& Trumper 1979) can be represented quite precisely by a  thermal
Comptonization spectral model.  It was thus shown that X-ray radiation in
the energy band from a few to several hundred keV is a result of the soft disk
photons upscattering off thermal electrons. Chakrabarti \& Titarchuk (1995
hereafter CT95) subsequently demonstrated through hydrodynamical calculations, a
centrifugal-barrier shock region is formed in the region around compact
objects (10-20 $R_s$) where the accreted matter releases its gravitational energy
and heats that  region to temperatures of 30-70 keV. In TLM98, the authors
argued that any Keplerian motion in  the disk has to pass through the centrifugal barrier
(CB) region if the inner disk boundary rotates with a sub-Keplerian velocity.
The CB location depends on the effective Reynold's number which, they  point out, leads to CB
oscillations which can in turn lead to an observable QPO signature.

The presence of the bulk motion in addition to thermal motion in the
Compton cloud is an unavoidable consequence of the gravitational
attraction by the central source. Blandford \& Payne (1981) were first to address the
issue of bulk-motion effects on the emergent spectra of accreting black holes by
formulating and solving the bulk motion problem in the simplified context
of a nonrelativistic Fokker-Plank treatment. They considered a semi-infinite
converging inflow atmosphere without taking into account any effects of
General Relativity. The first realization that the high-soft state black hole spectra
might be related to Bulk Motion Comptonization was put forth by Titarchuk, Mastichiadis \&
Kylafis [1996 and  in the extended version in (1997), hereafter TMK].

Zane et al. (1996) have elaborated the numerical method of characteristic
for the solution of general relativistic kinetic (GRK) equation for the bulk
inflow atmosphere. Subsequently, Titarchuk \& Zannias (1998 hereafter TZ98)  solved
semi$-$analytically the relativistic kinetic equation for a finite converging inflow
atmosphere using the separation variables method but they neglected the recoil effect in
photon-electron interactions. Laurent \& Titarchuk (1999)  reproduced TZ98's results through Monte
Carlo simulations and they calculated the overall spectrum taking into account all effects
of the photon-electron interactions and  the Special and General Relativity Theory.

Recently,  Papathanassiou \& Psaltis (2001, hereafter PP01) presented
calculations of the resulting spectrum of the bulk-motion atmosphere.
They solved numerically the general-relativistic kinetic (GRK) equation
in the steady-state Schwarzschild spacetime, neglecting the recoil effect
in photon-electron interactions. They demonstrated that power-law spectra result
from multiple scatterings, similar in some regards to thermal Comptonization.

 PP01 conclude that ``the photons emerging from the accretion flow with
high energies do not carry any signatures of regions of high velocities or space time".
However, their calculations do not allow them to make this strong
statement because they neglect the recoil effect. The estimate of the
mean electron energy  is determined by  the position of the high energy
cutoff in the spectrum. It is the same as in the thermal Comptonization
case (see for example ST80).  The spectral index itself which is a real outcome
of PP01 calculations does not allow for estimates of the average electron
energy.  {\it In TZ98 and PP01, following separate and independent
approaches to solving the same kinetic equation, both sets of authors elaborate
the necessary properties of the Bulk motion inflow in Schwarzschild space
time: spectral indices which converges to the asymptotics value of $\simeq1.7$,
and a hard-photon source distribution confined to $\sim3$ Schwarzschild
radii}.  For additional details, we refer the reader to the references
contained in  in TMK97, PP01 and Psaltis (2001).

 We also note recent work by Reig et al (2001), who have considered the
possibility that rotational motion of electrons in the disk upscatter photons to
produce the high-energy power law. They are similarly able to reproduce the basic
high-energy continuum properties, provided that the disk vertical optical depth
exceeds some threshold value near the inner-most Keplerian orbit.

\subsection{Predictions of Observable Black-Hole Properties}

Most of the spectral and temporal properties predicted theoretically by
the BMC model can in principle be verified by observations. They
are: ({\it i}) the specific high-energy spectral index, $\alpha\simeq 1.75$ in the limits
of high accretion rates and low electron temperatures of order 1 keV and
less; ({\it ii}) the presence of a high energy turnover, at energies between
300-500 keV;  ({\it iii}) QPOs detected as oscillations of the hard power
law component formed in the same inner-disk vicinity through bulk motion
upscattering. These are due to a hard photon-source distribution with a
strong maximum about (1.5-1.8)~$R_s$ (TZ98, Fig. 3, see also Laurent \&
Titarchuk 2001) which result from the
strong photon bending and Doppler effects in the immediate vicinity of a
black hole.  The contribution to the total disk flux from
the inner most region is generally small, hence these oscillations would
not be detected. ({\it iv}) a lack of discernible coherence between the
temporal signals of the high-state soft-thermal emission, emanating from a
relatively large region, and the hard-power-law emission emanating from a
compact region, and illuminated by a relatively small fraction of the
overall thermal emission,  and finally, ({\it v}) an increase in this coherence
measure as the source transitions to the low-hard state.

\subsection{Emergent Thermal Spectrum: The Hardening Factor}
Empirical support of the BMC model, based on numerous high-soft-state
BHC observations, has been presented in papers by Shrader \& Titarchuk
(1998, hereafter ShT98), Borozdin et al.(1999, hereafter BOR99) and
Shrader \& Titarchuk (1999, hereafter ShT99).  In BOR99 and ShT99 it was
shown through high quality fits of the BMC model to the observed spectra, that
the soft component can be extracted with a high degree of accuracy. The
emergent soft flux component depends strongly on BH mass, distance and a color
correction or "hardening factor" $T_h$, which is the ratio of the
effective to observed blackbody color temperature (see BOR99).
Because the mass and distance are known to a high degree of accuracy for
GRO J1655-40, it was possible to determine $T_h\approx 2.6$.

Recently  Zhang et al. (2000) and Merloni, Fabian \& Ross (2000)
have also pointed out the importance of the hardening factor determination.
There is an essential difference between the BOR99 empirical
determination and the Zhang et al (2000) determination.
In BOR99 the authors extracted the blackbody-like component from the
high-soft spectrum of GRO J1655-40 using BMC's model. The quality of fit was very
high with $\chi^2$ per degree of freedom of order unity
(see BOR99, Table 3 and Fig. 1).  They found that the presence of
a blackbody-like spectral  component  was consistent with the data.
Then, assuming the thermal component emanates from the disk, the Shakura-Sunyaev 1973
model (hereafter SS73) is used to determine $T_h\simeq2.6$.
Recently, the value $T_h=2.6$ for GRO J1655-40 was independently confirmed
by Ebisawa et al. (2001).

In contrast with our determination Zhang et al. (2000) assumed that the
disk component is embedded in a Comptonizing "haze", and is thus viewed through this
medium. The hardening factor they derive is due then
to upscattering of the disk soft photons in this medium, and it thus
depends strongly on the optical depth and temperature (see Sunyaev \&
Titarchuk 1980). This being the case, the hardening factor can have
any of a wide range of values (Zhang et al. found that $T_h=3-6$). We concur
that at times the "weather" in GRO J1655 may be such that
the disk is covered by this relatively hot haze and additional spectral components
may be needed to  fit the data (see BOR99). But very often the effective emission
area of the disk is directly observable and its intrinsic properties can
thus be determined.

Merloni et al. (2000), using the numerical calculations of the disk
spectral formation, confirm that the hardening factor can be noticeably higher
than 1.7 which is widely assumed in the literature following Shimura \& Takahara (1995).
Furthermore they found (see Table 1 of their paper)
that with certain combinations of model parameters the hard color factor
can be very close to the empirically determined value $T_h=2.6$. It is worth noting
that the theoretically determined value of $T_h$ is very sensitive to choice of model
parameters, hence making it very difficult to use for BH mass determination.
Given the large uncertainty in the $\alpha-$parameter
for realistic black-hole accretion disks, and because $T_h$ changes
drastically when $\alpha$ approaches unity [which can be the case at least for disks
in NS binaries (Titarchuk \& Osherovich 1999)], we have opted to proceed using our
empirically determined value.
 

\section{THE EMPIRICAL PICTURE}

\subsection {High-Energy Spectral Evolution}

To illustrate some of these ideas we  present our recent analysis of the
X-ray nova XTE J1550-564 (e.g. Remillard et al 1999; Sobczak et
al. 1999) which was observed by RXTE and CGRO.  This analysis is based on data obtained
by the CGRO OSSE and BATSE instruments, and primarily, the RXTE PCA and HEXTE
instruments.

The basic BMC model features can be seen in the context of an evolving
source, in this case for XTE~J1550-564 during the March 1999 time
frame. The composite hard-soft X-ray light curve (Figure 1) illustrates roughly
the evolution of that particular event--its rather gradual linear decay
from about 2.5 to 2 Crab, followed by an exponential decay in the 2-10 keV
band, with more erratic flaring behavior in the BATSE 20-100 keV band.

The spectra illustrated in Figure 2, derived from the RXTE observations
of about TJD 11220 (where TJD=JD-2450000), represent the {\it extreme high
soft-state} in XTE~J1550- 564. The flux is completely dominated by  thermal emission,
with a characteristic temperature of $kT\simeq 1$~keV.
The mass-accretion rate is high, as indicated
by the $\simeq$2.5 Crab flux (2-10 keV), and we speculate that the
geometrical configuration of the thermal source expands due to viscous
angular momentum loss in the inner disk.  In the extreme case, only a small
fraction of the thermal photons illuminate the bulk-in-flow site.
Hard photons are further suppressed due to
Compton down-scattering by the thermal plasma. As the accretion surge
subsides, the situation moderates.  This is reflected in the de-convolved
spectra; the hard power-law component is barely discernible in the upper
panel, and then slightly more pronounced in the lower panel. The parameters of the fit
(panels a and b): (T,~$\alpha,~f$)=(0.9,3.7,0.02) and (0.9,2.6,0.06),
where $T$ is the temperature of the injected soft photons, $\alpha$ is the
energy spectral index  and $f$ is the illumination factor.

As the mass-accretion further decreases and the system stabilizes, the
hard X-ray flux increases in proportion to the thermal source. The
resulting spectral form is the characteristic black-hole {\it high-soft state}
spectrum, illustrated in two stages of evolution here.
In the spectrum illustrated in panel (c), which is derived from RXTE
PCA and HEXTE observations on TJD~11236, the hard power law is becoming
more prominent.  This is also evident in the upturn in the BATSE Light Curve.
The next spectrum of the sequence, (d), is a composite from XTE and
OSSE covering about 3-300 keV. The OSSE integration spans TJD~11253-11260.
The hard power law is now very prominent out to $\sim200$~keV where it
becomes noise limited. This is reflected in the evolution in our inferred
parameters, (T,$\alpha,~f$)=(0.9,1.6,0.11), that is the illumination
fraction has now increased.
It must be noted however, that the OSSE spectrum presented
here represents an integration over
the 1-week viewing period, whereas the RXTE soft X-ray
spectra are typically $\sim~10^3$ s ``snapshots''.
This more than two orders of magnitude difference in the integration time
for the RXTE/PCA and OSSE bands clearly makes problematic the
identification of a high-energy spectral cutoff.
Such a feature cannot be reliably established, primarily due to the
noise-limited nature of the high-energy, but
additionally because of the drastically different  integration times of two these
instruments (in this particular case, the net background subtracted
count-rate for the coarsely binned 400-650 keV interval is in fact negative by about
1$\sigma$). 
The long time integrations are unavoidable for the high energy part of spectra
because of the very steep power law shape. But from the other hand they are  problematic 
for the high-soft state, where any changes in the illumination geometry or mass accretion
rate (of any sub-Keplerian or Keplerian portions of the flow) may lead to the
drastic spectral changes.
 \par
As the outburst continues to evolve
(Figure 2e \& f), the luminosity of the thermal
source decreases. The ambient plasma cloud can no longer be efficiently
Compton cooled, and the thermal luminosity is thus further diminished by
scattering. The resulting decrease in the thermal excess is clearly seen
in this sequence of de-convolved spectra (in either case, a detection is
made only up to  20 keV). This change is quantitatively reflected in the
derived parameters: T decreases (0.75 to 0.34 keV),
and {\it f}, which recall is related to the illumination parameter of
the bulk-motion in fall, increases (0.2 to  0.31), although the
power-law index is still consistent with the high-soft-state spectral
form.

\subsection{Spectral Characteristics Associated with QPO Behavior}

In Figure 3, we show how continuum models derived through fits to BH
X-ray nova XTE~J15550-564 (panel a) and 4U~1630-47 (panel b).
For XTE~J1550-564 data from the two epochs; October 1998 (solid
curve), when 185-Hz QPOs were  reported (e.g. Remillard et al. 1999); and
March 1999 when no prominent high-frequency QPOs are present.
The October 1998 case represents a higher luminosity state
than March 1999. The spectra are scaled
for illustrative purposes. It is worth noting the dramatic difference in
the proportional luminosity in the soft-to-hard components for the two cases.
This is reflected in the inferred values of the $f$ parameters, 0.04 and
0.6 for the two cases.  4U~1630-47 provides another, albeit less dramatic
example (panel b).

We interpret this as being due to an enchanced illumination of the
converging inflow site, $1-4~R_S$ for the hard/QPO case, which we
conjecture results from a pile up of the material -- i.e.
the formation of a standing wave -- in the inner disk region (TLM). This
is also reflected in the larger parameter $f$, the BMC illumination
factor, of our model (ShT98-99). We consider the
{\it luminosity and spectral dependence, specifically
the $f$-dependence, of the presence (absence) of the QPO as an
additional support for the BMC model}.

\subsection{Mass-distance estimate for XTE J1550-564}

We have previously developed and presented a method
for physical parameter estimation based
the derived BMC model parameters (ShT99; see also BOR99). The basic idea
is to infer the effective soft-emission
region area in terms of the distance (typically unknown) and the
previously mentioned hardening factor. The observable surface
area can then be applied to a determination of the black-hole mass
in terms of a specific disk model, typically that of
SS73 with modification to treat electron scattering.

Applying our method to XTE~J1550-564, we
obtain the distance to mass ratio $d/m=0.042\pm 0.004$, where $m$ in solar units and
$d$ is in units of 10 kps. A $\cos(i)=0.5$ as a cosine of the  inclination angle and the
hardening factor $T_h=2.6$ has been assumed in this calculation. This suggests a
large black-hole mass, perhaps $m\simeq10-15$ if the source is in the 5-kpc
distance range. The distance is unknown for  this source, but given the large
column density $N_H=2\times10^{22}$ it is probably more than a few kpc.  We
further note that for $d=5$ kpc and our  corresponding mass estimate, the source
would have been radiating in the 5-7\% $L_{\rm Edd}$ range during early March
1999 when the 2-10 keV flux was in  the 2-2.5 Crab. However, given the
apparently large mass accretion rate 10-15\% $L_{\rm Edd}$ is more plausible and
the distance may well be in the $7-8$  kpc range calling for a larger BH
mass.

\section{Discussion}

\subsection{Spectral/Temporal Connections in BH Binaries}

There are roughly speaking three ``types" of QPO behavior seen in Galactic
black hole binaries; $(\sim .1,~10,~100)$ Hz. The latter 2 seem to be
associated with "power law-dominant" high-soft states
(e.g. Remillard. et al 1999; Cui et al 1997).
The high-frequency  cases are also luminosity dependent -- they tend to
be associated with luminosity surges -- perhaps for the same reasons that
kilohertz QPOs in  neutron star sources are associated with high
luminosity states of those objects. The high-frequency QPO amplitude is also seen
to increase with energy [see e.g. Morgan et al (1997); Cui et al (2000)].
In our interpretation, this is because the high-energy photons are
produced in a more compact region -- i.e. the converging inflow site -- than the
low-energy ones which are produced in the inner-disk annuli.
Another empirical trend that has emerged is that the
break frequency, $\nu_b$, of the power
density spectra (PDS) increases with the photon energy (Belloni et
al. 1997, Homan et al. 2000).
Titarchuk \& Osherovich  (1999) have shown that the break frequency
$\nu_b$ is related to the size of the emission region. This is because $\nu_b$ is
inversely related to the emission area viscous time $t_b$.
Thus smaller values of $t_b$ imply smaller emission areas for a given
energy band. This is precisely the case for Bulk Motion Comptonization;
harder photons are produced in the deeper layers of the converging inflow
atmosphere.

We should also note that the coherence between low  and high energies,
e.g. between  $\sim$ 2-5~keV and $>6$~keV, is small, in the soft-high
state but it increases  when the soft-to-hard  state transition occurs (Cui et
al. 1997; Nowak et al. 2000).  In the framework of the BMC model, only the
innermost part of the disk, of order a few $R_S$, provides the seed
photons for the Bulk motion Comptonization and thus for hard photon
production. The low degree of coherence between intensity fluctuations of the low-energy
disk photons ($\sim$ 2-5 keV), which come from much larger disk area of order $\pi
[(10-15)R_S]^2$, with that of the high energy photons formed in the
compact converging inflow region of order $\pi (3 R_S)^2$ arises
naturally. However the coherence is expected to increase for the soft-hard transition because
the Comptonization region becomes larger  and ultimately  it completely covers
the seed-photon area of the disk.

 Each of these trends  is {\it difficult to explain in context of
``standard" disk plus coronae models}. They can in principle be reproduced
by invoking additional parameter, such as a radial temperature gradient in
the corona (e.g. Nobili et al. 2000; Lehr, Wagoner, \&  Wilms  2000), or by more elaborate
geometrical and dynamical configurations (e.g. Boettcher \& Liang 2000). However, we
assert that {\it they are explained naturally and self consistently if the hard
flux emanates from a compact region}. In this case, the: (i).
high-frequency inner-disk oscillations are expected to be preferentially
scattered, since only this  small fraction of the thermal source
illuminates CI region, (ii). higher energy photons are produced closer to the
BH  horizon (where the velocity is closer to the speed of light), and
thus the compactness of the emission region increases with the energy,
(iii). high luminosity is consistent with ``pile up" in accretion flow.

\subsection{Hard and Soft Phase Lags}
Additional important information related to the X-ray spectral-energy
distribution can be extracted from the time lags between different energy bands.
While hard lags have been observed for several sources, soft phase lags
have been found in some cases as well (e.g. Reig et al. 2000).
In addition, there have been suggestions of separate harmonic components of
given feature having lags of the opposite
signs [e.g.  Tomsick \& Kaaret (2000) studied the QPO properties of
GRS 1915+105 where both negative (soft) phase lags and positive (hard) lags
were detected]. However, {\it it is natural to expect the positive time lags in the
case of the thermal Comptonization. In the case of bulk-motion Comptonization,
both negative and positive time lags are anticipated}.
 In the thermal Comptonization case, the primary soft photons gain energy in
 process of scattering off hot electrons; thus the hard photons spend more time
in the cloud  than the soft ones. There is then a one-to-one correspondence
between the photon energy and the photon travel time in the cloud.
But this is not the case for bulk-motion Comptonization where soft disk
photons at first gain energy in  the deep layers of the converging inflow,
and then in their subsequent path towards the observer lose
energy in the relatively cold outer atmosphere. If the overall optical
depth of the converging inflow atmosphere (or the mass accretion rate) is near
 unity, we would detect only the positive lags as in the thermal
 Comptonization case, because relatively few photons would lose energy in
 escaping. But with an increase of the
optical depth, the soft lags appear because more hard photons lose their
energy in the cold outer layers. There is a piling up effect at low energies
due to the recoil effect (see ST80).
A significant portion of the low-energy photon distribution,
particularly at energies of order 2-5~keV, results from this
piling up effect. The low-energy photons spend more time in the matter than the
higher-energy photons because since they undergo additional scatterings from
cold electrons of the outer regions.

In fact, for unsaturated Comptonization the spectral formation
is   mostly due to the Doppler effect but in the saturated case it is equally
determined by recoil and the Doppler effects (see e.g. ST80 and figures of spectra there).
 For thermal Comptonization
the relative change of energy $\Delta E/E=(4kT_e-E)/m_ec^2$. In the unsaturated case,
the photon energy E, on average, is less than $4kT_e$ and thus  most of the photons gain
energy and the soft photons are led by the hard ones and thus the hard lags occur
due to Comptonization. But in the saturated case when the Comptonization parameter
is very high, i.e. when the product of $4kT_e/m_ec^2$ and $\tau^2$ is greater than one,
a significant fraction of the hard photons lose energy trying to thermalize to the
average electron energy, $ kT_e$.  In this case there is a distribution of
soft and hard lags as a result of Comptonization. Some of hard photons are followed by
the soft ones.

A very similar situation is realized for the Bulk Comptonization where the temperature
of the electrons should be replaced by $m_e v_b^2/3$ (TMK97, appendix D). 
Because in  the outer part
of the converging inflow atmosphere
$(r>5r_s)$  $m_e v_b^2/3=m_ec^2(r_S/r)/3$ less than 30 keV the hard photons with energies
more than 30 keV readily lose their energies and consequently the hard photons can be
followed by the photons with energies less than 30 keV.

Because the hard state is characterized by the thermal Comptonization
spectra we expect the hard lags  in the hard state and
{\it we expect the soft lags in the high-soft states characterized by
high {\it f} values} (i.e. power-law dominated).
Hard lags would be more likely to occur in high-soft states with lower
{\it f} values.
Laurent \& Titarchuk (2001) present the extensive calculations of the
distribution of time lags in the framework of the BMC model.

\subsection{Additional Spectral Components}
We also note that we have not included additional continuum features
such as Compton reflection in our high-soft state analysis (e.g. Tomsick
et al. 1999; Zdziarski, 2000; Gierlinski et al. 1999). While it may always be
possible to improve the chi-square statistic through the  introduction of
additional parameters, our goal is to extract physical information
invoking a minimal parameter space.
Furthermore, the aforementioned analyses were motivated by the presence
of significant residuals to additive continuum models (typically blackbody
or multi-color disk plus power law). In our fitting however, we have not
generally been so compelled. For example, for the GRO~J1655-40 high-soft
state covered by the TJD~10322-10328 OSSE observation we obtain a 3-parameter
fit with a reduced chi-square statistic of $\chi^2_\nu\simeq 0.6$. This compares
favorably to other efforts in the literature where more complex models are invoked.
For example, a $\chi^2_\nu\simeq 0.7$ result was obtained  by Tomsick et al (1999)
by invoking a 8-parameter model including Compton reflection characterized
by a 26\% covering factor. The difference may have to do with the effects of
applying an additive model as opposed to a self-consistently derived model
to the instrumental response (Figure 4).

\subsection{Extent of the High-Energy Power Law}
Another issue regards the high-energy extent
of the hard power law. As noted, the BMC model predicts a
turnover of this power law below the electron rest-mass energy
due to Compton recoil effects, relativistic bending of photon trajectories
and gravitational redshift. There have been claims in
the literature of detections to as high as 700 keV, however, what is generally
presented is the high-energy extension of a power-law $fit$ to the data, which is in fact
dominated by the higher signal-to-noise detector count rates at lower
energies. The problem is that the
source count rates above $\sim500$~keV are typically a fraction of a
percent of the background count rates, and the long integration times required
lead to systematic uncertainties which are hard to characterize.
The first, and still the most notable case has been GRO~J1655-40
(e.g. Grove et al 1997; Grove et al. 1998; Tomsick et al 1999).
Referring to our analysis of GRO~J1655-40 from the
preceding section, a coarse binning of the data leads to a statistical
significance of $\sim$3.5-$\sigma$ in the 475-600~keV spectral range. To
better address this issue we have analyzed 14 spectra from 6 Galactic
BH binaries observed with OSSE during the CGRO mission. The resulting
statistical distribution of detection significance for the
$\simeq450-750$~keV band is illustrated in Figure 5. Although, there
is an apparent positive asymmetry relative to the overlayed
Gaussian curve (which represents random statistical
fluctuation about a zero mean), the results seem less than compelling,
given the likely predominance of systematic uncertainties in this
spectral domain.

We further note that while we have made a specific prediction regarding
the high-energy extent of the BMC-produced high-energy power law,
we are not ruling out the possibility of higher-energy gamma-ray
emission from some other process.

The recent claim by Zdziarski et al. (2001) regarding the observational evidence of the nonthermal
Comptonization and ruling out the bulk Comptonization for the high-soft state is based on the
OSSE observation of the high-soft state in GRS 1915+105. The exposure time for these observations
of $(2-7)\times10^{5}$ s is incomparable with the dynamical time scale of the X-ray emission region
of order of $10^3$ s determined by the subKeplerian component.
One should expect variability of the spectral index of the emergent spectrum within one hour or
less. Furthemore, Focke et al. (1997) analyzed RXTE observations of Cyg X-1 in 
the soft state.  They  claimed   significant variations of the hard tail 
photon index  occuring in minute time scales. 
 Thus it is very problematic to restore the true spectral shape (particularly the high energy
turnover)  using the data with the  exposure time of order of a few times of $10^5$ seconds when
the real count rate from the source  is washed out and additionally distorted by the high level
X-ray background.

\subsection{Sub-Keplerian Flows}
CT95 first argued that  two flows (subKepleian and Keplerian) are always
present in the binaries. Now this prediction is confirmed by the recent observation of a number
of sources including XTE J1550-554 (Smith et al. 2001; Soria et al. 2001)
The dynamical time scale of  the propagation of the subKeplerian components towards the central
object  is very close to the free-fall time scale which is less that $10^3$ s for a black hole
of 10 solar masses.

\subsection{Physical Parameter Extraction: Previous Approaches}
We note that many efforts described in the literature have attempted
to directly infer physical  information through spectral deconvolutions
based on specific disk models. The models are typically some variation of the
SS73 formulation. We point out however, that the SS73 formulation invokes
a dimensionless radial integration variable, and thus for example,
{\it determinations of the inner disk radius in physical
units are not valid}.

A full analytical presentation of the multi-color disk spectrum
can be found in ShT99, equations 1 and 2. In this analysis,
the spectrum is formulated as an integral over disk annuli
with a lower limit expressed as the dimensionless inner radius,
namely, $r_{\rm in}=R_{\rm in}/R_{\rm S}$. Here $R_{\rm S}$
is the Schwarzchild radius. The temperature distribution in
the SS73 formulation is described by a function $f(r)$, where
r is the dimensionless radius in Schwarzchild units
(ShT99, equation 2). Mitsuda et al. (1984) [also see
Makishima et al. (1986)] first suggested
simplification of this integral (e.g. ShT99 equation 1)
by ignoring the term $(r_{\rm in}/r)^{1/2}$.
{\it This simplification  transforms the radial integration over the disk
from dimensionless to dimensional units. Such a procedure is not mathematically
correct}. It is nonetheless widely used in the community; it is for example, the
formalism underlying the ``diskbb'' routine of the ``XSPEC'' package.

Despite this problem, in many instances in the literature
of this field, values for the inner-disk radius are not only quoted
in physical units to high degrees of precision, but the accompanying
dialog often leaves the reader with the misleading impression that
what is being presented is a {\it measurement},
rather than an {\it inferred parameter of a particular (and
non-unique) spectral deconvolution}.

\subsection{Global Disk Oscillations, Low Frequency QPOs and Mass
Determination}
The low frequency oscillations seen in hard state as well as soft, and
at multiple wavelengths have been interpreted as the  global disk mode
(GDM) oscillations (Titarchuk \& Osherovich, hereafter TO00).  If the disk
oscillates coherently as a single body, the observed frequencies should
be seen in all wavelengths emitted by the extended disk, i.e. from optical
to X-rays. The GDM frequency,  of order  0.1 Hz for  BH masses of order of
10-20 solar, is consistent with the observations (TO00). This lack of
energy dependence, as well as the observed anticorrelation of the QPO
frequency with the disk mass has been observed in XTE~J1118+480
(Wood et al. 2000). The latter effect is also a natural  consequence of
GDM model, where QPO frequency anticorrelates with the disk mass which is
believed to decrease during the soft-to-hard transition. The
intensities and  profiles of high-excitation optical/UV lines, believed
to result from photoionization of the outer disk by X-irradiation,  may
also be modulated at the GDM  frequency.
This is a measurement which could in principle be made as a further
test, and it also allows for a more precise estimates of the size and
physical properties of the disk (Wood et al 2001).

The low-frequency QPOs, interpreted in the context of the GDM
oscillations and the BMC model analysis provide two independent
derivations of BH masses.
This approach  has been applied to derive the BH mass in LMC X-1
(TO00) which was earlier derived using the X-ray spectral method (ShT99). A similar
procedure can be applied to  XTE~J1550-56, XTE~J1859+227, and others.
Recently McClintock et al. (2000); also see Wagner et al (2001) have independently confirmed
earlier, more speculative results presented by Dubus et al. (2000) regarding the
binary solution for XTE~J1118+480. Those authors demonstrated
through observations of the quiescent
system that the velocity amplitude of the dwarf secondary is $698 \pm14 $km/s  and the
orbital period of the system is $0.17013 \pm 0.00010$ d. The implied value of
the mass function, $f(M) = 6.00 \pm 0.36$ solar
masses, provides a hard lower limit on the mass of the compact primary that
greatly exceeds the maximum allowed mass of a neutron star. Among the eleven
dynamically established black-hole X-ray novae, the
large mass function of XTE J1118+480 is rivaled only by V404 Cygni and XTE~J1859+226.
A large black hole mass, in XTE J1118+480 in fact, was
predicted earlier by TO00  using the GDM frequency. The GDM leads to a BH mass estimate
for XTE J1118+480 of $\sim 7M_{\odot}$ based on the $\sim 0.1$ Hz QPO frequency, in agreement with
the optical observations by Wagner et al. ( We note there is a typographical error in equation (14)
of TO00 which should have $x_{\rm in}^{-4/5}$ instead of $x_{\rm in}^{-8/15}$.)
 We note that distance-to-mass determination method based on the BMC
formalism (BOR99, ShT99 and section 4.5) could not be applied to this source since it observed
only in the low-hard  spectral state.

\subsection{High - Frequency QPOs}
As a synopsis of the QPO behavior in BH binaries it is worthwhile to
note that sources such as XTE J1550-56, 4U 1630-47, XTE J1859+227
exhibit $\sim 10^2$  Hz QPO behavior only when  high ``{\it f}" values are seen in our analysis
({\it f} parameterizes the fractional illumination of the CI region; see
BOR99). The intermediate case, i.e. $\sim 10$~Hz QPOs, are more ambiguous and the
low-frequency  ($\sim10^{-1}$) QPOs seem to be energy independent,
spanning the X-ray to near IR bands (Haswell et al. 2000; Wood et al 2000;
Chaty et al 2000). Low-frequency QPOs are seen in low-hard and high-soft
states but high-frequency cases only in high-soft states (Remillard et al 1999; Cui
1998). Thus, there is  a apparently certain connection between different
QPO regimes. The observations seem to be suggesting that  the presence of the low
frequencies is not associated  with spectral state transitions, whereas
high frequencies appear only in the high-soft state, albeit an extreme
manifestation (hard-power-law dominated) of that state.
It is worth noting that in GRS~1915+105, a broad PDS
feature near 70 Hz is detected the low-hard state whereas a
strong QPO feature is observed in the high-soft state (Trudolyubov
2000). In our interpretation, the oscillations in the near vicinity of the black hole
are strongly smeared because they are viewed through a hot Compton cloud
by the the observer. This prediction, based on the converging-inflow
conceptual frame work should be confirmed or refuted by observations.

\section{CONCLUSIONS }

 Based on a number of empirical trends in the literature regarding the
temporal and spectral properties of BH binaries, supplemented by our own
analysis of a number of these objects, we have presented a number of
arguments in support of the BMC model. Most notably, the high-soft state spectral form,
and the high frequency variability but long-term stability of the power
law, which are problematic for thermal Comptonization scenarios, arise
naturally in our model. These features, along with the other empirical trends such as
the lack of temporal coherence between the soft and hard variability in the
high-soft state, and increased coherence in the low-hard state further
support our conjecture of the existence of a compact high-energy emission
region located in the vicinity of BH horizon $(1-4R_S)$. The total
(kinetic) energy of the matter increases towards to the BH horizon.
This radial gradient of the matter energy  provides a natural explanation of the noted
increase in QPO power with energy. A majority of efforts to explain these
phenomena invoke thermal-Comptonizing media external to the
disk. The problem with this approach, from our point of view, is that the
physical parameters of the Comptonizing plasma require "fine tuning" to
retain consistency with the ubiquitous nature of the spectral form. This is
further compounded by the requirement to provide a self consistent explanation for the
recently emerging temporal characteristics, such as this energy
dependence of the high-frequency QPO power. If one for example, invokes
radial thermal gradients to a Comptonizing plasma as an explanation of the
QPO energy dependence, the inferred parameter space must be further tuned
to allow one to consistently reproduce the emergent high-soft state
spectral form over a considerable luminosity range.

Regarding the high-frequency QPOs, we predict analogous effects in AGN
but with frequency rescaled in inverse proportion to BH mass, thus at
least seven orders of magnitude lower, of order of $10^{-5}$ Hz. In particular,
the narrow-line Seyfert-1 Galaxies which are
an characterized by the steep power law with $\alpha\sim 1.7$ and 
 apparent soft excess emission, relative to
the canonical $\alpha\sim0.8$ X-ray spectra of the broader class of Seyfert-1
and quasar spectra. This may be related to mass-accretion rate (and thus
to the host environment) in a manner similar to the bimodal spectral behavior
of Galactic binaries, thus these objects would be the obvious candidates
to search for this phenomenon. Nominal evidence for such oscillations has
recently appeared in the literature (Boller, et al 2001).

We argue that the high-soft-state spectral form represents a unique
black-hole event horizon signature. We further conjecture that an
extra-Galactic analog may exist, namely the narrow line Seyfert-1 galactic
nuclei. Given the ubiquitous nature of this phenomenon for many Galactic
systems, and its apparent extension into the extra-Galactic realm, we
suggest that we are witnessing a direct signature of the disappearance of
matter falling down the black-hole event-horizon drain.

\vspace{0.2in}

\centerline{\bf{ ACKNOWLEDGMENTS}}
This work made use of the High-Energy Astrophysics Science Archive
Research Center, the Compton Gamma Ray Observatory and Rossi X-Ray
Timing Explorer science support facilities at the NASA Goddard Space
Flight Center.

\begin{figure}
\epsscale{0.7}
\plotone{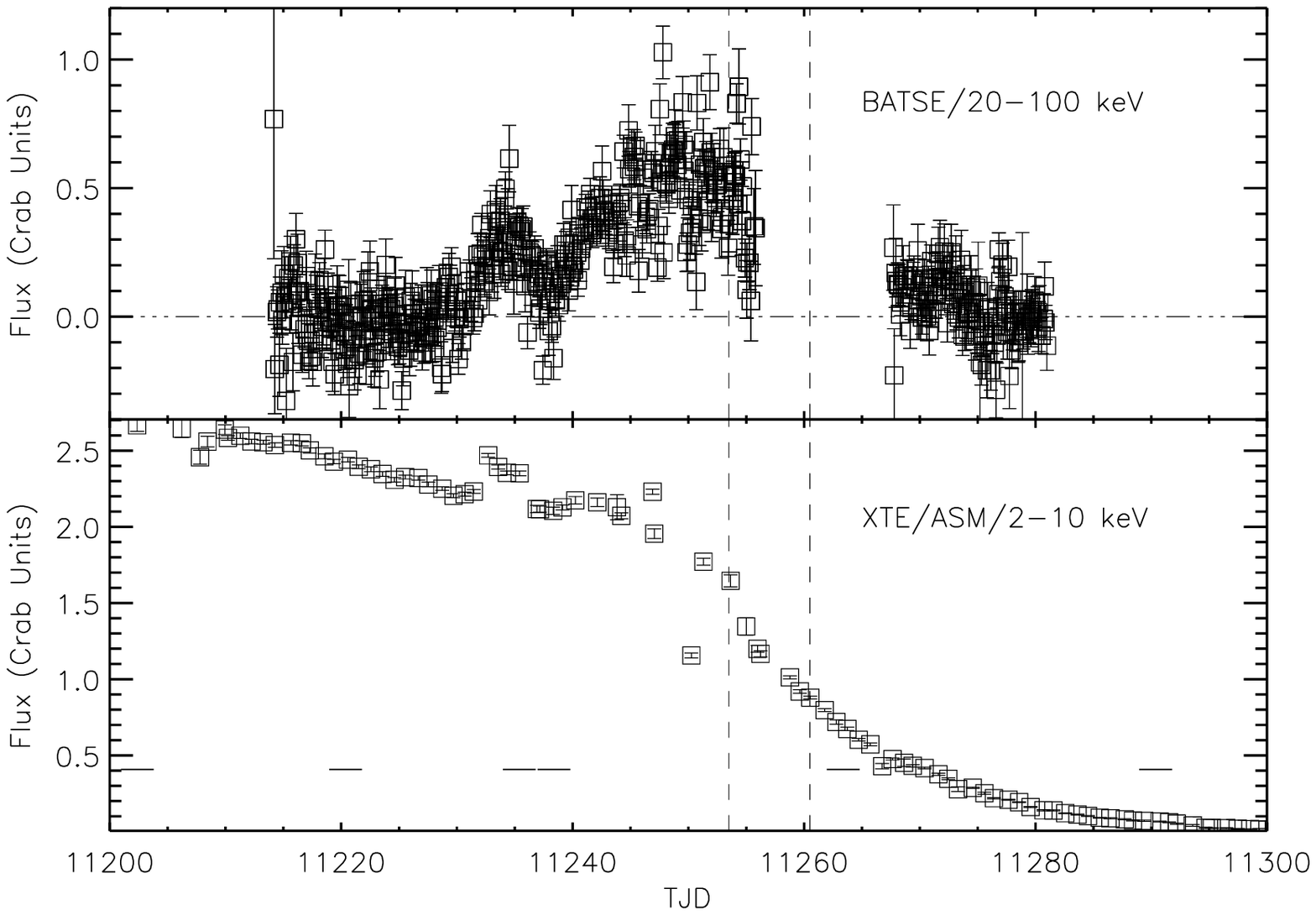}
\caption{CGRO/BATSE and RXTE/ASM light curves covering the Spring 1999 active
period of XTE J1550-564. The vertical, red, dashed lines indicate the VP
808.5 OSSE observation, and the short horizontal lines indicate pointed
RXTE observations used in our analysis.  The hard X-ray flux seems to be
declining sharply during the OSSE observation (there is a gap in the BATSE
coverage).}
\label{fig1}
\end{figure}


\begin{figure}
\includegraphics[width=2.7in]{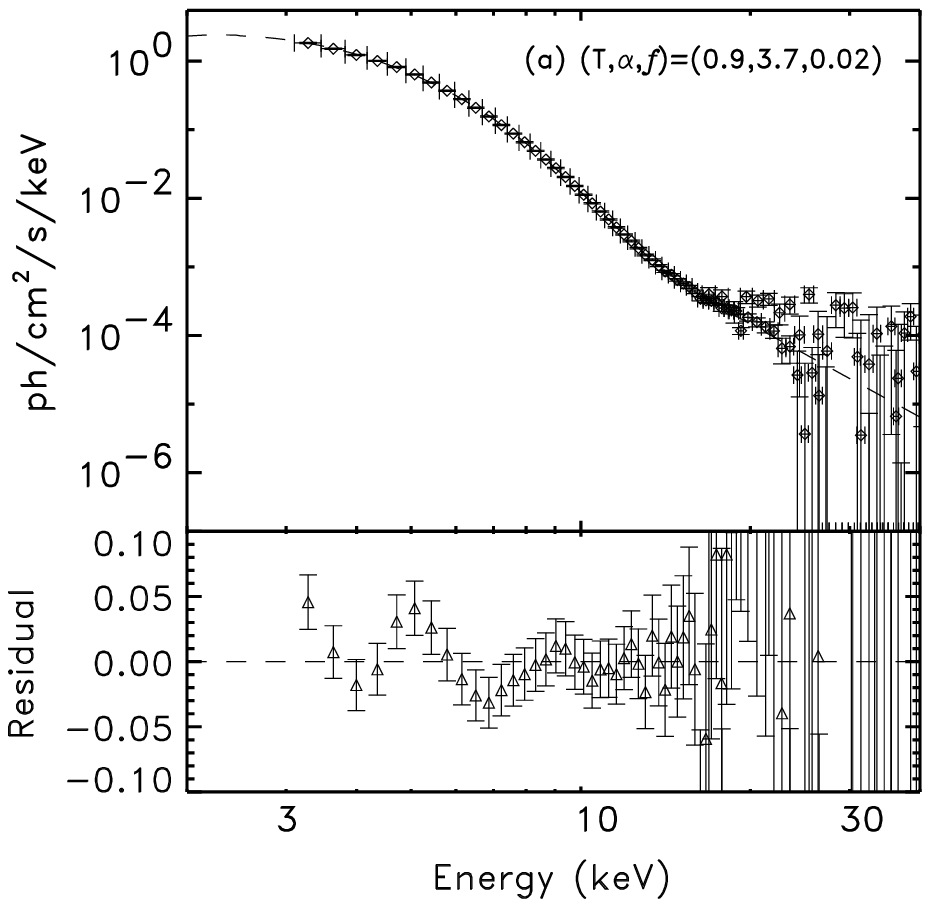}\includegraphics[width=2.6in]{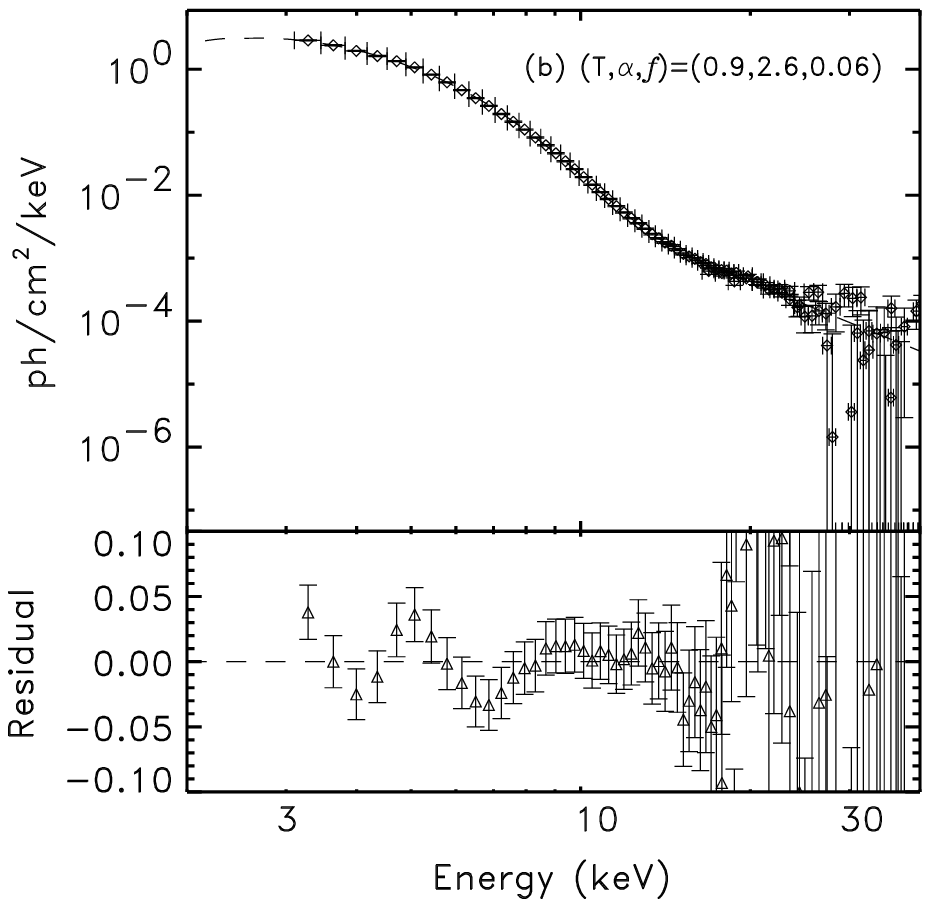}
\includegraphics[width=2.5in]{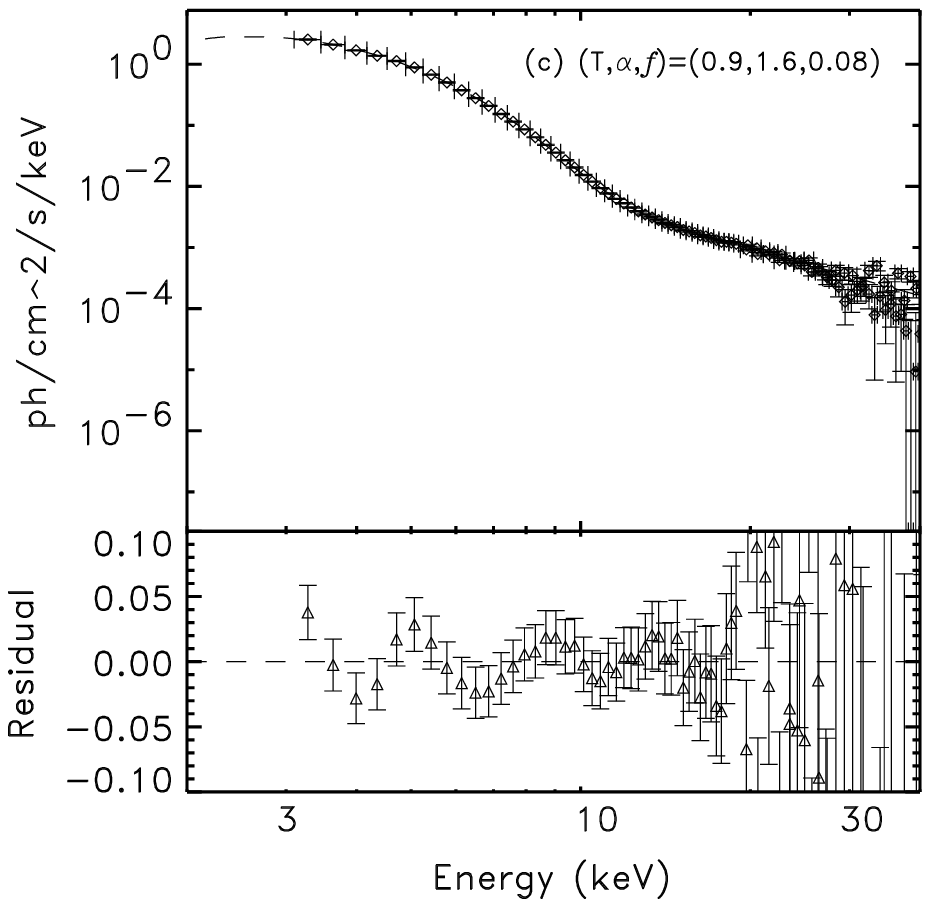}\includegraphics[width=2.5in]{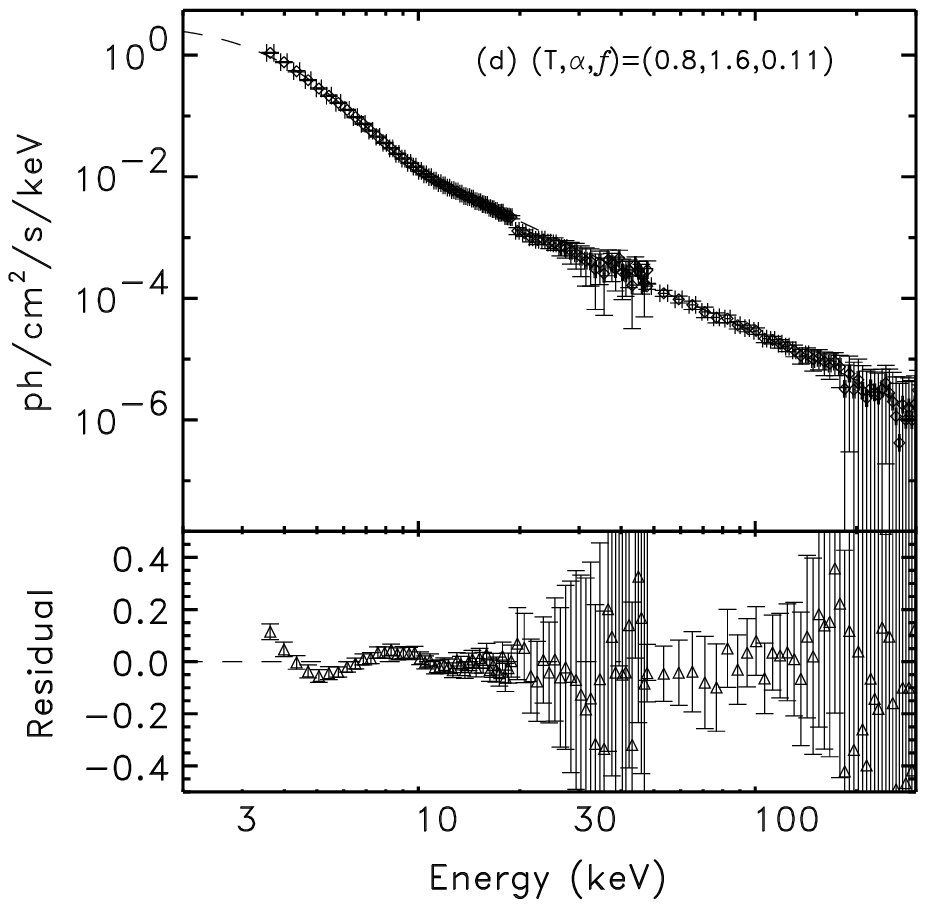}
\includegraphics[width=2.5in]{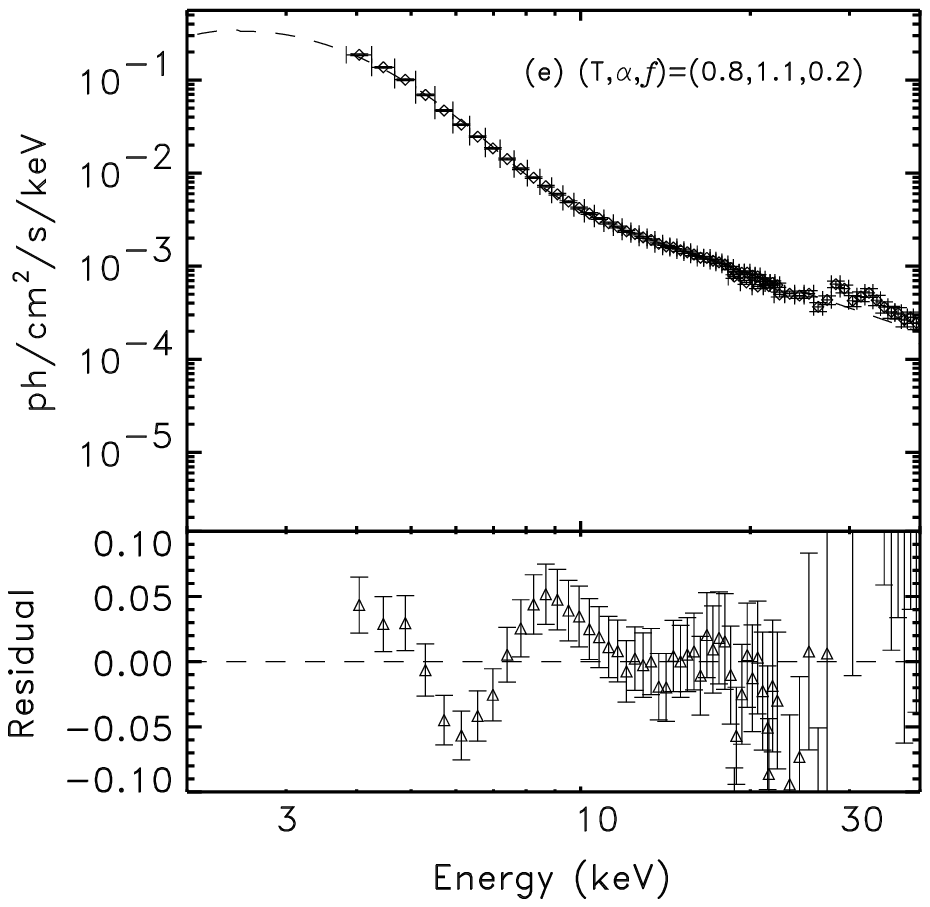}\includegraphics[width=2.5in]{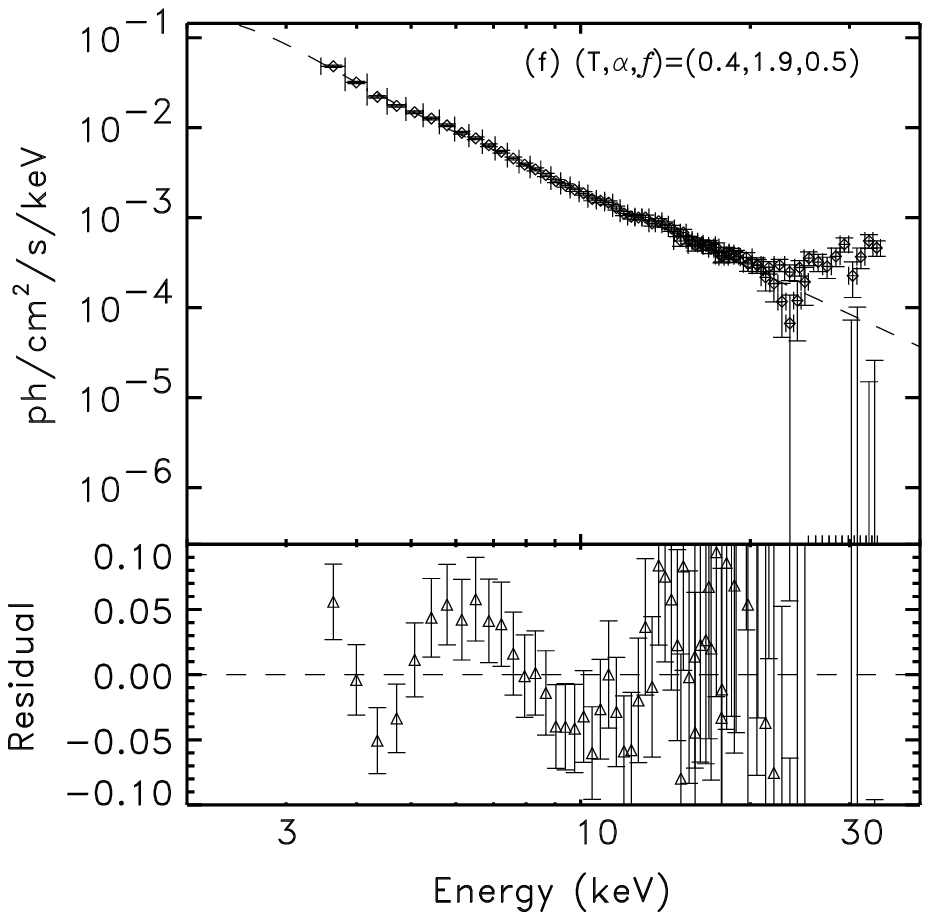}
\caption{Evolution of the high-energy spectral energy distribution in
XTE~J1550-564. (a) represents the extreme
high-soft state (about TJD~11220) when the hard-power law is marginal or
absent.  (b) is similar to (a) but the hard-power law is beginning
to become more prominent. Panel (c) a transition to the more familiar high-soft
state has occurred. Panel (d) is similar to (c). 
In panel (e)  (TJD~11262) the luminosity has decreased significantly and 
the transition into the low-hard state is evident. 
In panel (f) (TJD~11293) is similar to (e).  
\protect\label{fig2}}
\end{figure}

\begin{figure}
\includegraphics[width=3.3in]{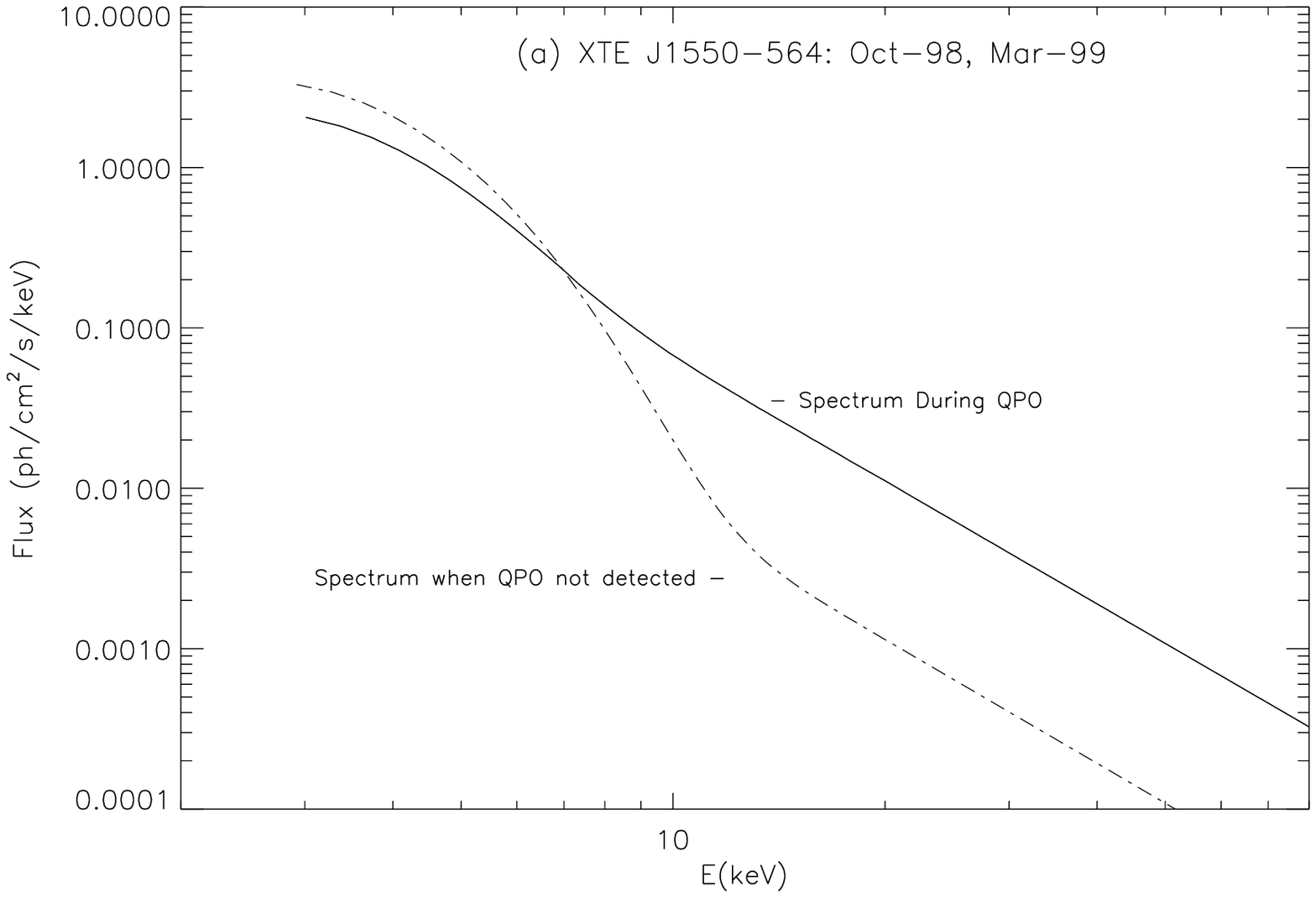}\includegraphics[width=3.3in]{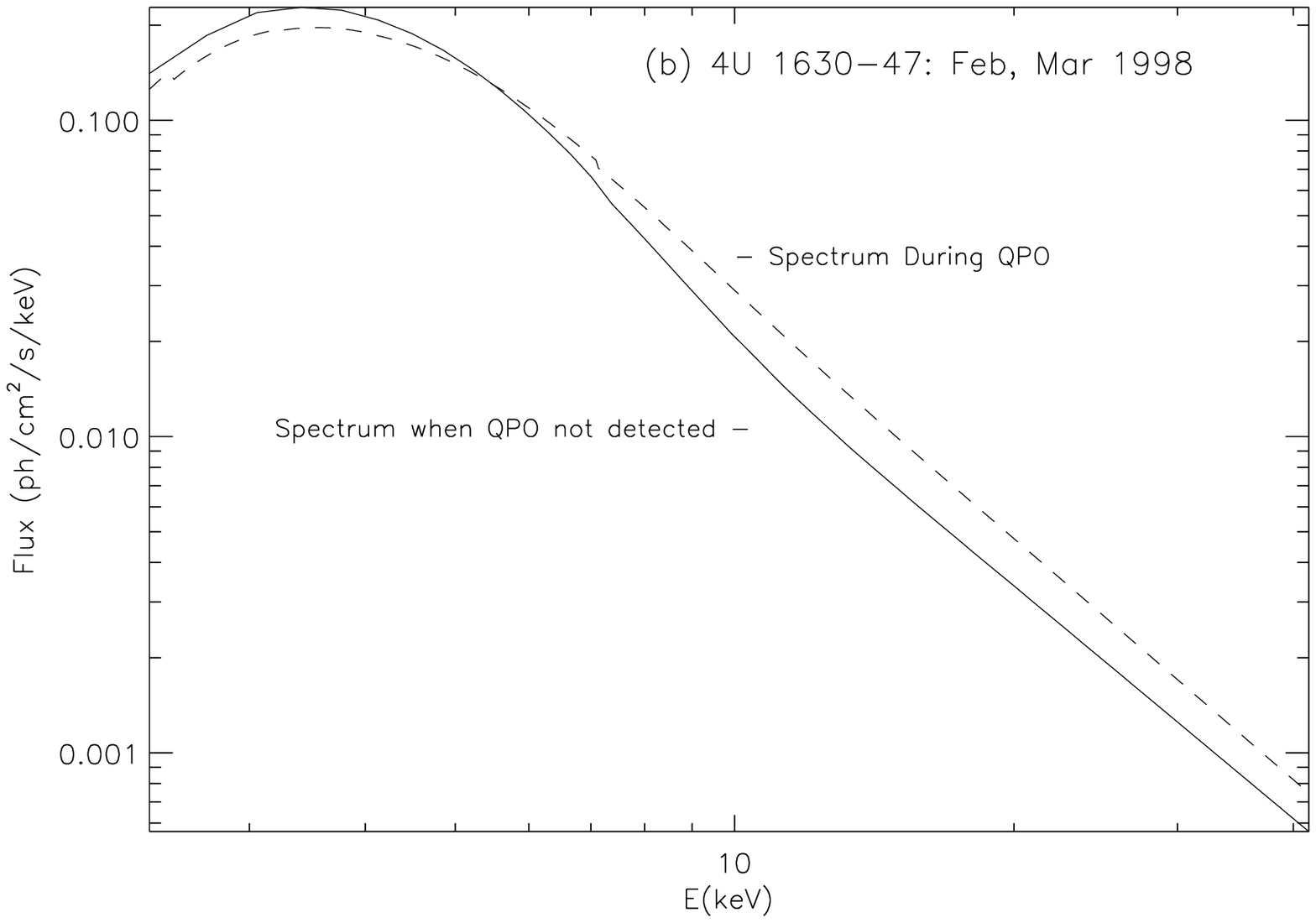}
\caption{ Continuum models fitted to the BHC X-ray nova (a) XTE J1550-564 at two
epochs: October  1998 (solid curve), when 185-Hz QPOs were reported and
March 1999 (dashed line) when no QPOs were observed.
The relative normalization has been adjusted for illustrative purposes
(the October 1998 case corresponds to higher flux state than March 1999).
 Note the distinctly different proportions
of the hard - power law and soft thermal components. We interpret this as
differences in the illumination geometry of the converging inflow site,
parameterized by $f$ in our model. 
Continuum models fitted to the BHC X-ray nova.
In (b) similar, although less dramatic,
effects are seen in 4U~1630-47 at two (QPO and non-QPO) epochs. 
\protect\label{fig3}}
\end{figure}

\begin{figure}
\epsscale{0.7}
\plotone{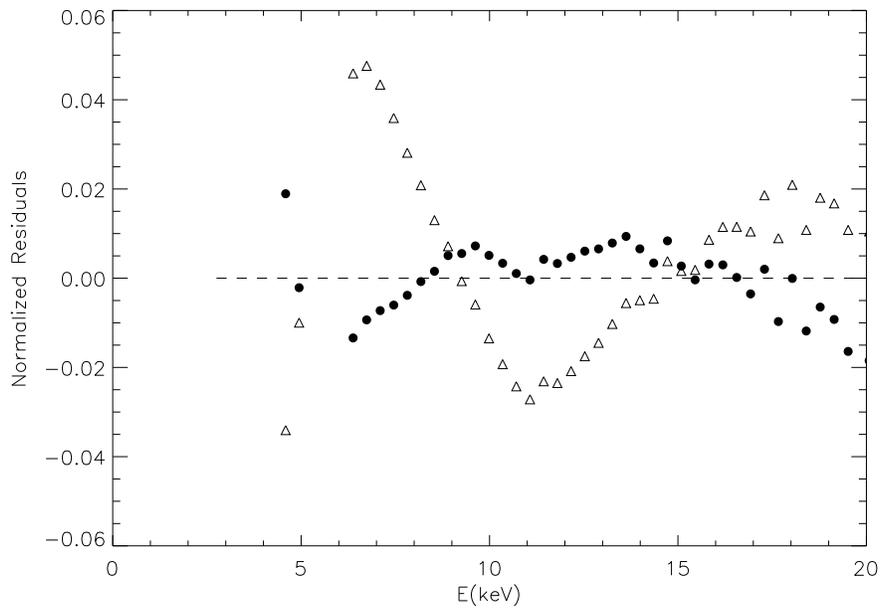}
\caption{Normalized residuals (data minus model) for BMC (filled circles) and
additive power law plus black-body disk (triangles). Note that while the BMC
residuals are essentially contained within the several percent  level,
the additive model deviates significantly between 6 and 7 keV, and  above
about 15 keV. Such residuals could be misinterpreted as real spectral
features. \protect\label{fig4}}
\end{figure}


\begin{figure}
\epsscale{0.7}
\plotone{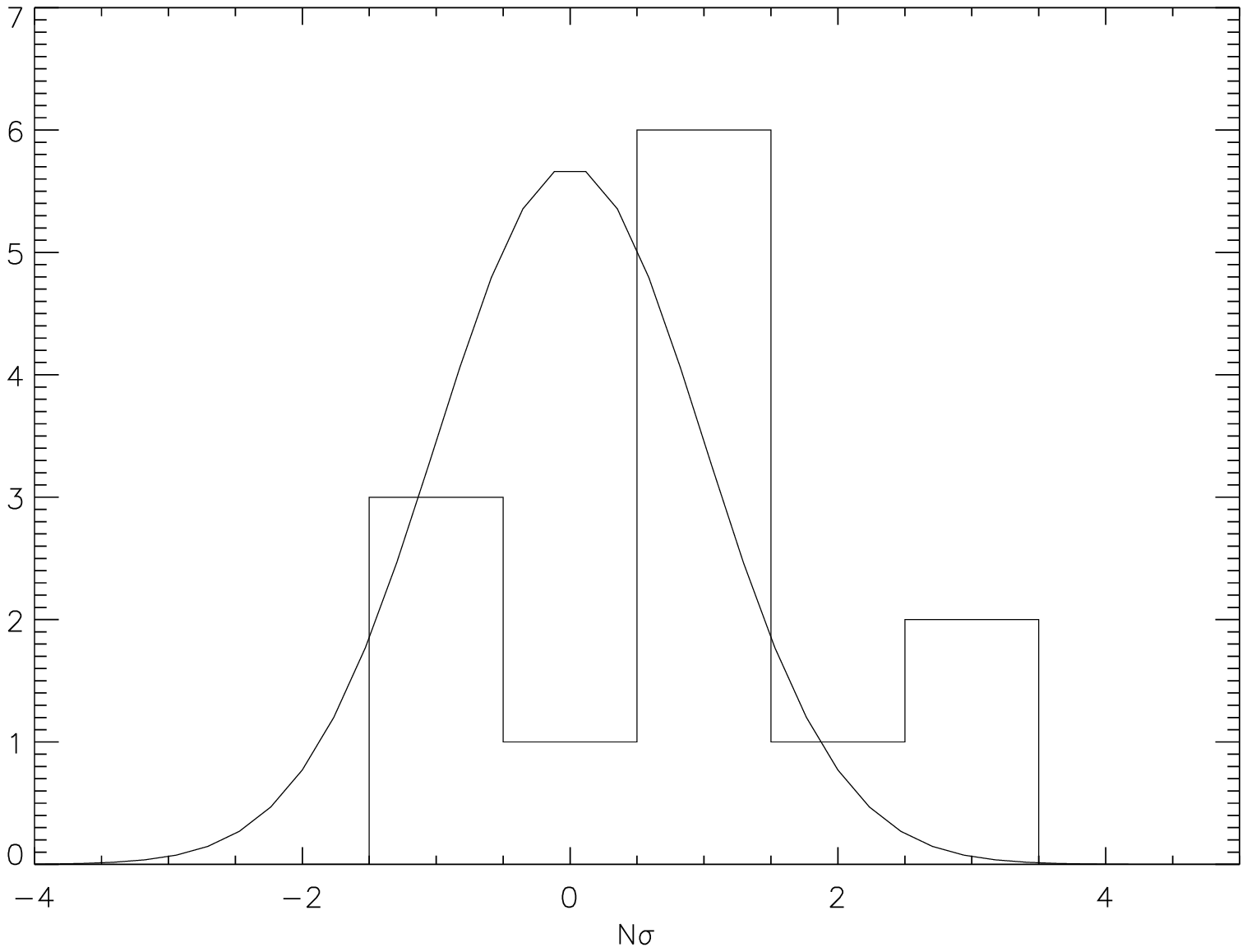}
\caption{Statistical distribution of the $\sim$450-750~keV detection
significance resulting from our analysis of 14 spectra from
6 Galactic BH binaries observed with OSSE during the Compton GRO mission.
The overlying curve, is a Gaussian with zero mean normalized to the data distribution.
\protect\label{fig5}}
\end{figure}






\newpage
\begin{thebibliography}{}
\bibitem[]{}
 Barret, D., \& Grindlay, J.E.  1995, ApJ, 440, 841

\bibitem[]{}
Blandford, R.D., \& Payne, D.G. 1981, MNRAS, 194, 1033

\bibitem[]{}
Belloni, T et al. 1997, A\&A, 322, 857

\bibitem[]{}
Boller, T.  et al. 2001, A\&A, 365, L146 

\bibitem[]{}
 Borozdin, K., Revnivtsev, M., Trudolyubov, S., Shrader, C.R.,
\& Titarchuk, L.G. 1999, ApJ, 517, 367 (BOR99)

\bibitem[]{}
Bottcher, M. \& Liang, E.P. 2000, ApJ, submitted, astro-ph/0003139

\bibitem[]{}
 Burderi, L., Di Salvo, T., La Barbera, A., \& Robba, N.R.
2001, ApJ, in press (astro-ph/0009183)

\bibitem[Chakrabarti \& Titarchuk (1995)]{ct95}
Chakrabarti S.K. \& Titarchuk, L. G. 1995, ApJ,  455, 623

\bibitem[]{}
Chaty, S., et al. 2000, In proceeding of Integral workshop, Alicante,
Spain 

\bibitem[Chen et al. (1997)]{csl97}
 Chen, W., Shrader, C.R., \& Livio, M.  1997, ApJ, 491, 312

\bibitem[]{}
Cheng, F.H., Horne, K., Panagia, N., Shrader, C.R., Gilmozi, R.,
Paresce, F., \& Lund, N. 1992, ApJ, 397, 664

\bibitem[]{}
Cui, W., Shrader, C.R., Haswell,C.A., \& Hynes, R.I. 2000, ApJ, 535,
L123

\bibitem[]{}
Cui, W., Zhang, S.N., \& Chen, W. 2000, ApJ, 531, L45

\bibitem[]{}
 Cui, W.,  Zhang, S. N., Focke, W., \& Swank, J. H. 1997,  ApJ,
v.484, p.383

\bibitem[]{}
 Di Salvo, T., Stella, L.,  Robba, N.R., van der Klis, M., Burderi, L.,
Campana, S., Frontera, F., Israel, G.L., Homan, J. \& Parmar, A.N.
2000, ApJ, 544, L119

\bibitem[]{}
 Dubus, G., Kim, R. S. J.,  Menou, K., Szkody, P. \&
Bowen, D. V. 2001, ApJ, 553, 307





\bibitem[]{}
Focke, W., Swank, J., Phlips, B., Heindl, W. \& Cui, W.  1997, 
In proc. 4th Compton Symposium, AIP-CP410  (Eds. C.D. Dermer, 
M.S. Strickman , \& J.D. Kurfess) 2,   854


\bibitem[]{}
 Gierlinski, M., Zdziarski, A.A, Poutanen, J., Coppi, P., Ebisawa, K.
 \& Johnson, W. N. 1999, MNRAS, 309, 496.

\bibitem[]{}
Grove, J.E., Johnson, W.N., Kroeger, R.A., McNaron-Brown, K., \& Skibo, J.G.
 1998, ApJ, 500, 899 

\bibitem[]{}
Grove, J.E. et al., 1997, In proc. 4th Compton Symposuim, AIP-CP410, 
(Eds. C.D. Dermer, M.S. Strickman \&  J.D. Kurfess) 1, 122

\bibitem[]{}
Haswell, C.A., Skillman, D., Patterson, J., Hynes, R.I., \& Cui, W.
2000, IAU, Cirl. 7427

\bibitem[]{}
Heindl, M., \& Smith, D.M. 1998, ApJ, 506, L35

\bibitem[]{}
Homan, J., Wijnands, R., van der Klis, M., Belloni, T., van Paradijs,
J., Klein-Wolt, M., Fender, R., \& Mendez, M. 2001, ApJS, 132, 377


\bibitem[]{}
Hua, X-M., \& Titarchuk, L.G. 1995, ApJ, 449, 188

\bibitem[]{}
Iaria, R., Burderi, L., Di Salvo, T., La Barbera, A., Robba, N.R.
2001, ApJ, 547,412

\bibitem[Laurent \& Titarchuk  (2001) ]{lt01}
 Laurent, Ph., \& Titarchuk, L.G. 2001, ApJ, 562, L

\bibitem[Laurent \& Titarchuk  (1999) ]{lt99}
 Laurent, Ph., \& Titarchuk, L.G. 1999, ApJ, 511, 289  (LT99)

\bibitem[]{}
Liang, E.P. 1998, Phys. Rep., 302, 66.

\bibitem[]{}
Makishima, K., et al., 1986, ApJ, 308, 635

\bibitem[]{}
Merloni, A., Fabian, A.C., \& Ross, R.R. 2000, MNRAS, 313, 193

\bibitem[]{}
McClintock, J.E., Garcia, M.R., Caldwell, N., Falco, E.E., Garnavich, P.M.
\& Zhao, P. 2001, ApJ, 551, L147

\bibitem[]{}
Mitsuda, K., et al., 1984, PASJ, 36, 741

\bibitem[]{}
Morgan, E.H., Remillard, R.A., Greiner, J., 1997, ApJ, 482, 993

\bibitem[]{}
Nobili, L., Turolla, R., Zampieri, L., \& Belloni, T. 2000, ApJ, 538, L137

\bibitem[]{}
Nowak, M.A., Wilms, J., Heindl, W.A., Pottschmidt, K., Dove, J.B. \&
Begelman, M.C. 2001, MNRAS, 320, 327 




\bibitem[]{}
Papathanassiou, H., \& Psaltis, D. 2001 MNRAS, submitted
(astro-ph/0011447)

\bibitem[]{}
Poutanen, J. \& Svensson, R. 1996, ApJ, 470, 249

\bibitem[]{}
Psaltis, D. 2001, ApJ, 555, 786 


\bibitem[]{}
Reig, P., Kylafis, N., \& Spruit, H.C., 2001, A\&A, 375, 155

\bibitem[]{}
Remillard, R.A., McClintock, J.E., Sobczak, G.J., Bailyn, C.D., Orosz,
J.A., Morgan, E.H., \& Levine, A.M. 1999, 517, L127

\bibitem[Shakura \& Sunyaev (1973)]{SS73}
 Shakura, N.I., \& Sunyaev, R.A. 1973,  A\&A, 24, 337 (SS73)

\bibitem [Shimura \& Takahara (1995)]{shimura95}
 Shimura, T., \& Takahara, F. 1995, ApJ, 445, 780

\bibitem[]{}
 Shrader, C.R., \& Titarchuk,L.G. 1999, ApJ, 521, L121 (ShT99)

\bibitem[]{}
 Shrader, C.R., \& Titarchuk,L.G. 1998, ApJ, 499, L31 (ShT98)

\bibitem[]{}
 Smith, D.A., Heindl, M., Markwardt, C.B., \& Swank, J.H.  2001, ApJ, 554, L41

\bibitem[]{}
Sobczak, G.J., McClintock, J.E., Remillard, R.A., Levine, A.M., Morgan,
E.H., Bailyn, C.D., \& Orosz, J.A. 1999, 517, L121

\bibitem[]{}
 Soria, R., Wu, K., Hannikainen, D., McCollough, M. \& Hunstead, R.  
2001, A\&A submitted astro-ph/0108084

\bibitem[]{}
Strickman, M., \& Barret, D. 1999, In proc. of the fifth Compton
Symposium, AIP-CP510 (Eds. M.L. McConnel and J.M. Ryan) p. 222


\bibitem[]{}
Sunyaev, R.A., \& Titarchuk, L.G. 1980, A\&A, 86, 121

\bibitem[]{}
Sunyaev, R.A., \& Titarchuk, L.G. 1978, Preprint of the Space Research
Institute of Soviet Academy of Science, 441.

\bibitem[]{}
Sunyaev, R.A., \& Trumper, J. 1979, Nature, 279, 506

\bibitem[]{}
Titarchuk, L., \& Osherovich, V. 2000a, ApJ, 537, L37 (T000a)

\bibitem[]{}
Titarchuk, L., \& Osherovich, V. 2000b, ApJ, 542, L111  (T000b)

\bibitem[]{}
 Titarchuk, L., \& Osherovich, V. 1999, ApJ, 518, L95 (T099)

\bibitem[Titarchuk \& Zannias (1998)]{TZ98}
 Titarchuk, L., \& Zannias. T. 1998, ApJ, 493, 863 (TZ98)

\bibitem[Titarchuk, Mastichiadis \&  Kylafis, (1997)]{tmk97}
 Titarchuk, L. G., Mastichiadis, A.,  Kylafis, N. D. 1997, ApJ, 487, 834
(TMK)

\bibitem[Titarchuk, Mastichiadis \&  Kylafis, (1996)]{tmk96}
 Titarchuk, L. G., Mastichiadis, A.,  Kylafis, N. D. 1996, A\&AS, 120,
C171

\bibitem[]{}
Tomsick, J.A., Kaaret, P., Kroeger, R.A., \& Remillard R.A. 1999,
ApJ, 512, 892

\bibitem[]{}
Tomsick, J. A., \&  Kaaret, Ph.  2001, ApJ, 548, 401
(astro-ph/0009354)

\bibitem[]{}
Trudolyubov, S.P. 2001, ApJ, 558, 276



\bibitem[]{}
Wood, K. S., Titarchuk, L.G., Ray, P.S, Wolf, M.T., Lovellete, M.N.
\& Bandyaopadhyay, R.M.  2001,  ApJ, 563, 

\bibitem[]{}
Wood, K. S. et al. 2000, ApJ, 544, L45 

\bibitem[]{}
Zane, S., Turolla, R., Nobili, L., \& Erna, M. 1996, ApJ, 466, 871

\bibitem[]{}
Zdziarski, A. A., Grove, J.E., Poutanen, J., Rao, A.R., \& Vadawale, S.V. 2001, ApJ, 554, L45
L45

\bibitem[]{}
Zdziarski, A. 2000, IAU Symposium (Eds. P.C.H. Martens, S. Tsuruta \&
M.A. Weber), 195, p. 153 , v3, astro-ph/0001078

\bibitem[]{}
Zhang, S.N., Cui, W., \& Chen, W. 1997, ApJ, 482, L155

\bibitem[]{}
Zhang, S.N., Cue, W., Chen, W., Yao, Y., Zhang, X., Sun, X., Wu, X-B.,
\& Xu, H. 2000, Science, 287, 1239

\bibitem[]{}
Zycki, P., Done,C., \& Smith, D. 1998, ApJ, 496, L25.

\end{thebibliography}
\end{document}